\documentclass[sigconf,10pt,screen]{acmart}
\AtBeginDocument{%
  }

\acmYear{2025}\copyrightyear{2025}
\setcopyright{rightsretained}
\acmConference[HOTOS 25]{Workshop in Hot Topics in Operating Systems}{May 14--16, 2025}{Banff, AB, Canada}
\acmBooktitle{Workshop in Hot Topics in Operating Systems (HOTOS 25), May 14--16, 2025, Banff, AB, Canada}
\acmDOI{10.1145/3713082.3730374}
\acmISBN{979-8-4007-1475-7/25/05}

\usepackage[utf8]{inputenc}
\usepackage{booktabs}
\usepackage{cleveref}
\usepackage{xcolor}
\usepackage{outlines}
\usepackage{caption}
\usepackage{subcaption}
\usepackage{paralist}
\usepackage{xspace}
\usepackage{siunitx}
\usepackage{enumitem}

\newcommand{\grumbler}[2]{}
\newcommand{\changebars}[2]{#2}
\newcommand{\pagelimit}[1]{}
\newcommand{\todo}[1]{}
\newcommand{\souj}[1]{}
\newcommand{\nc}[1]{}
\newcommand{\raf}[1]{}
\newcommand{\mka}[1]{}
\newcommand{\neil}[1]{}

\crefname{section}{\S}{\S\S}
\crefformat{section}{\S#2#1#3}

\newcommand{\eg}{{\it e.g.,}\xspace}
\newcommand{\ie}{{\it i.e.,}\xspace}
\newcommand{\allnotes}[1]{}

\newcommand{\eat}[1]{}

\begin{document}

\makeatletter
\def\blfootnote{\xdef\@thefnmark{}\@footnotetext}
\makeatother

\date{} 

\title{Real Life Is Uncertain. Consensus Should Be Too!}




\author{Reginald Frank\textsuperscript{*}}
\email{reginaldfrank77@berkeley.edu}
\orcid{0000-0002-0423-1071}
\affiliation{%
  \institution{UC Berkeley}
  \city{Berkeley}
  \state{CA}
  \country{USA}
}

\author{Soujanya Ponnapalli\textsuperscript{*}}
\email{soujanya@berkeley.edu}
\orcid{0009-0006-1449-1447}
\affiliation{%
  \institution{UC Berkeley}
  \city{Berkeley}
  \state{CA}
  \country{USA}
}

\author{Octavio Lomeli}
\email{octavio.lomeli@berkeley.edu}
\orcid{0009-0006-8258-627X}
\affiliation{%
  \institution{UC Berkeley}
  \city{Berkeley}
  \state{CA}
  \country{USA}
}

\author{Neil Giridharan}
\email{giridhn@berkeley.edu}
\orcid{0009-0008-2175-2611}
\affiliation{%
  \institution{UC Berkeley}
  \city{Berkeley}
  \state{CA}
  \country{USA}
}

\author{Marcos K. Aguilera}
\email{mkaguilera@gmail.com}
\orcid{0000-0003-3489-2468}
\affiliation{%
  \institution{Broadcom}
  \city{Palo Alto}
  \state{CA}
  \country{USA}
}

\author{Natacha Crooks}
\email{ncrooks@berkeley.edu}
\orcid{0000-0002-3567-801X}
\affiliation{%
  \institution{UC Berkeley}
  \city{Berkeley}
  \state{CA}
  \country{USA}
}

\renewcommand{\shortauthors}{Reginald Frank\textsuperscript{*}, Soujanya Ponnapalli\textsuperscript{*}, Octavio Lomeli,\\ Neil Giridharan,
 Marcos K. Aguilera, and Natacha Crooks}

\begin{abstract}
Modern distributed systems rely on consensus protocols to build a fault-tolerant-core upon which they can build applications. Consensus protocols are correct under a specific failure model, where up to $f$ machines can fail. 
We argue that this $f$-threshold failure model oversimplifies the
real world and limits potential opportunities to optimize for cost or performance. We argue instead for a probabilistic failure model that captures the complex and nuanced nature of faults observed in practice.
Probabilistic consensus protocols can explicitly leverage individual machine \textit{failure curves} and explore side-stepping traditional bottlenecks such as majority quorum intersection,  enabling systems that are more reliable, efficient, cost-effective, and sustainable.

\end{abstract}



\maketitle

\blfootnote{These authors co-led the project and contributed equally}

\section{Introduction}
\label{sec:introduction}


Modern distributed systems rely on a \textit{fault tolerant core} that provides the abstraction of a single 
failure-free component atop which application-logic is implemented. At scale, failures are the norm, not the exception.
Most cloud-native databases~\cite{corbett2013spanner, taft2020cockroachdb}, configuration managers~\cite{hunt2010zookeeper, kubernetes}, decentralized platforms~\cite{ccf, ia-ccf, signal, ethereum}, and the latest AI model training and serving platforms~\cite{shen2024hugginggpt} build on fault-tolerant cores with consensus protocols~\cite{lamport2001paxos, ongaro2015raft, p-pbft, junqueira2011zab, giridharan2024autobahn} as their key building block.
Consensus protocols guarantee data reliability under a specific \textit{fault model}. A fault model captures one's belief about the world;
it defines the specific assumptions made by the protocol. In the case of consensus, developers must make
assumptions about the network, the type of faults, and the number of faults. The choice of fault model is
crucial as it significantly impacts protocol design and performance. Much like car insurance, an overly optimistic fault model will reduce the cost of consensus but may not capture reality, causing the system to fail. At one extreme, one can, for example, consider a fault model where no faults are considered possible! In contrast, an overly pessimistic fault model may unnecessarily complicate the protocol. 

Correctly capturing the reality of faults is a complex task. As such, most systems today simplify reality and guarantee correctness under the \textit{$f$-threshold} model. The system is safe (all nodes agree on the committed data) and live (all new operations are eventually committed) up to $f$ node faults. Nodes can fail either by crashing (requiring \textit{crash fault tolerance} or CFT) or through malicious compromise where nodes can deviate arbitrarily from the protocol (requiring \textit{Byzantine fault tolerance} or BFT).
 If more than $f$ nodes fail, the system provides no guarantees. Reconfiguration can progressively replace faulty nodes by correct ones to ensure that the fault threshold is never exceeded. 
 
This abstraction is clean but too simplistic: it hides important aspects of faults in modern systems and aligns poorly with how people reason about reliability today.
\par \textbf{Faults are complex.} The $f$-threshold model classifies servers as correct or faulty, and it treats faults uniformly: each fault contributes one unit to the current fault count,
  without recognizing that some servers are more prone to faults than others.
In reality, all servers eventually fail and each server has a unique probability of failing that depends on its type~\cite{backblazeBackblazeDrive} or even its location in the datacenter~\cite{seemakhupt2023cloud}. This probability changes over time as a function of device age~\cite{backblazeBackblazeDrive,dixit2021silent}, update rollouts~\cite{crowdstrike}, or workload shifts~\cite{yang2022spot}. Moreover, server
faults are neither uniform nor independent, as faults tend to cluster around software rollouts, unexpected workload shifts, or when new vulnerabilities are discovered~\cite{googlesecurityreport,chen2019sgxpectre,fei2021security,li2022systematic}.
\par \textbf{Guarantees are only probabilistic.} Current consensus protocols report all-or-nothing guarantees: they claim to be fully safe and live with fewer than $f$ faults, but provide no guarantees otherwise. This is unrealistic. As faults are probabilistic, it is \textit{always} possible for the number of faults to exceed $f$. Thus no consensus protocol can offer a guarantee stronger than probabilistic safety or liveness. Storage systems recognise this reality already. S3, for instance, describes its guarantees~\cite{s3dur} in terms of \textit{nines} of availability (99.9\%) and durability (99.999999999\%).

In this paper, we argue that consensus protocols should acknowledge reality:
in real deployments, all servers may fail, and do so with different likelihood. As such, no consensus protocol is 100\% safe or live. In fact, we find that Raft, a popular CFT protocol, is only 99.97\% safe and live in three node deployments when nodes suffer a 1\% failure rate (\S\ref{sec:analysis}). In our opinion, a consensus protocol should offer a specific safety (respectively liveness) \textit{probabilistic} guarantee. This guarantee should be computed as a function of the specific protocol, and of servers' specific \textit{fault curves}. Fault curves capture the unique, time-dependent, fault profile of a given server and can be computed using the large amount of telemetry that modern deployments track on a daily basis.

Moreover, adopting a probabilistic approach to reasoning about failure allows us to investigate an intriguing opportunity: that of leveraging the heterogeneous fault curves of different servers to provide the same probabilistic safety/liveness guarantees but at much lower dollar or energy cost! We find, for instance, that one can run Raft on nine, less reliable nodes that suffer a 8\% failure rate and obtain the same 99.97\% safety and liveness. If these resources are 10$\times$ cheaper (\eg spot instances~\cite{microsoftGeneralAvailability}, older hardware), this yields a 3$\times$ reduction in cost. This analysis hinges on incorporating fault curves in existing consensus protocols. We believe that one can go a step further and develop \textit{probability-native} consensus protocols, which use fault curves to
improve performance and reduce cost, by side-stepping expensive quorum intersection invariants that are essential to traditional consensus.

In the rest of this paper, we first describe the properties of real-world faults (\S\ref{sec:failures}).  We then conduct a detailed analysis of the probabilistic safety and liveness guarantees (\S\ref{sec:analysis}) of well-known CFT and BFT protocols~\cite{ongaro2015raft,castro2002practical}. 
We conclude by asking: what are the challenges and opportunities of designing true probability-native consensus protocols (\S\ref{sec:call}). 


\section{Faults are probabilistic}
\label{sec:failures}

The $f$-threshold model discretizes reality in the name of simplicity: it assumes that at most $f$ nodes are faulty, where correct nodes \textit{never} fail. It further does not recognize that some servers are more likely to fail than others, as
faults are treated equally: they each contribute one unit to the current fault count.
%
We argue that this simplification is counter-productive. It hides the true nature of faults and leads to consensus protocols that 1) offer guarantees that align poorly with how people think of reliability, and 2) miss opportunities for performance optimizations. Server faults should instead be modeled as probability distributions to account for their inherent heterogeneity and dynamic nature. We refer to this as a server's \textit{fault curve} $p_u$.

The extensive research on hardware faults at scale~\cite{hochschild2021cores,dixit2021silent,frigo2020trrespass,greenan2010mean,xu2019lessons} can help us precisely characterize $p_u$. Large-scale datacenters keep detailed telemetry on the observed fault rates of their servers~\cite{dixit2021silent} or GPUs~\cite{kokolis2024revisiting}, including the number of memory faults, bit flips or block corruptions on disk drives, and how they vary based on manufacturer, write cycles, or time. Similar work exists on quantifying spot instances failure rates~\cite{yang2022spot,amazonawsSpotInstance,microsoftAzureSpot}. In addition to experimental results, advanced analytical studies also predict fault rates. Advanced predictive models, for instance, can estimate fault rates based on transistor and chip aging~\cite{agarwal2007circuit,anik2020effect,amrouch2015connecting,amrouch2015techniques}. Large organizations similarly conduct large-scale threat and risk analysis to capture the likelihood of machine compromise and attacks~\cite{sommerstad2010probabilistic}.
 
Large-scale fault studies draw three primary conclusions: \\
\textit{(1) Nodes do not fail equally}. Most consensus protocols assume all nodes are equally likely to fail. In reality, fault probability is rarely homogeneous and depends on the hardware manufacturer,
~\cite{hochschild2021cores,pinheiro2007failuretrends,backblazeBackblazeDrive}, where the hardware is placed in the datacenter~\cite{seemakhupt2023cloud}, disk capacity or usage~\cite{pinheiro2007failuretrends,seemakhupt2023cloud}. Probability of failure is not limited to machine faults but can additionally be used to capture social concepts like human trust and incentives. In the distributed trust context, a lower fault probability can be assigned to parties with whom a long-standing contractual agreement exists~\cite{mazieres2015stellar}, whereas members from an enemy state may be assigned a higher fault probability. Stake in blockchain systems captures a similar idea: nodes with higher stake have more to lose if the system fails, and thus are considered more trustworthy~\cite{algorand,ethereum}.
\par \textit{(2) Fault likelihood evolves over time.}
Consensus protocols do not currently reason about fault likelihood, and can thus only react to evolving fault rates by changing the threshold $f$.  Unfortunately, changing $f$ is cumbersome as it requires costly reconfiguration~\cite{duan2022dyno}. Yet, fault probabilities evolve over time. At the software level, faults tend to cluster around major software updates as seen with the CrowdStrike debacle~\cite{crowdstrike}, or with peak operation hours and sudden workload changes. Hardware reliability also evolves ~\cite{dixit2021silent,hochschild2021cores}. 
Disk failures, for instance, follow a bathtub curve: they have a high chance of failure during the \textit{infancy} and \textit{wear-out stage}, but comparatively lower failure rates during the \textit{useful life stage}. Both Google and Facebook report that silent corruption errors (processors that compute data incorrectly) become more frequent as cores age~\cite{dixit2021silent,hochschild2021cores}.  In the context of distributed trust, fault probability (aka trustworthiness) may evolve as a function of the geopolitical context.

\par \textit{(3) Faults are correlated.}  Protocols assume a maximum of $f$ faults and treat faults uniformly, implicitly assuming that faults are independent.
Unfortunately, faults often are correlated or planned. At the software level, they arise from periodic reboots, software rollouts, or operational updates. At the hardware level,
research shows that devices used in a similar way, or placed close to each other exhibit similar fault patterns. For instance, disks that share similar vibration or temperature exhibit similar fault patterns~\cite{seemakhupt2023cloud}. Malicious attacks also frequently compromise \textit{classes} of machines. Consider for instance the distributed trust setting such as Azure Confidential Ledger or Signal's key recovery service. Both use trusted hardware like Intel SGX or AMD SEV-SNP to strengthen defenses against a machine compromise. When a vulnerability is discovered in those architectures, as is sadly frequent~\cite{googlesecurityreport,qiu2019voltjockey,chen2019sgxpectre,fei2021security,li2022systematic,li2021cipherleaks}, the risk of platform-wide attacks grow.
\par \textit{(4) Faults cannot be simply treated as crashes or Byzantine.}
Current consensus protocols,
with few exceptions~\cite{clement09upright} force developers into a stark choice: either optimistically assume that malicious faults will \textit{never} arise, or always pay the cost that they will. In reality, most nodes fail by crashing but from time to time exhibit malicious behavior. Consider for example the \textit{corruption execution errors} at Google and Facebook triggered by mercurial cores~\cite{hochschild2021cores,dixit2021silent}; these errors amount to Byzantine failures. They are, however, much rarer (approx. 0.01\% at Google) than traditional server faults (4\% Annual Failure Rate). The same rationale holds in the distributed trust setting: TEEs prevent Byzantine attacks most of the time, but undiagnosed vulnerabilities can lurk.


Reliability research in storage systems, unlike consensus protocols, has designed effective metrics to capture these patterns. Disk reliability is expressed in terms of Annual Failure Rate (AFR), often measured across a large fleet of disks~\cite{backblazeBackblazeDrive}.
The storage community relies on Markov models of their system to quantify metrics like Mean Time to Failure (MTTF), Mean Time Between Failures (MTBF), and Mean Time to Data Loss (MTTDL)~\cite{heart, patterson1988case}. In a Markov model, states capture configurations (\ie number of operational disks) and transitions resulting from disk failures, repair or recovery, with rates governed by failure probabilities ($\lambda$) and repair probabilities ($\mu$).
With steady-state probabilities, the expected values for MTTDL and MTTF guide the design of effective mechanisms for reliable systems (\eg the expected MTTDL with a striping scheme of $n$ disks and $k$ parity disks striping in RAID). These systems provide configurable \textit{nines} of probabilistic guarantees to applications, in line with how people reason about reliability today.
This rich knowledge of failures has been used to deploy erasure coded data in the cloud~\cite{kadekodi2020disk,heart}, obtaining over 40\% in disk cost savings. 

We believe that a similar approach is possible in consensus. We foresee two approaches: 1) better understand and exploit the probabilistic guarantees offered by \textit{existing} consensus protocols, and 2) revisit whether the core primitives of consensus (quorum intersection, leader election, etc.) can be redesigned to use fault curves and probabilistic guarantees. We describe each in turn. 

\section{Analysis of Consensus Protocols}
\label{sec:analysis}

We first analyze existing consensus protocols to understand what guarantees they offer when thinking of faults as probabilistic. For simplification, we do not consider reconfiguration
(adding or removing nodes) and assume faults are independent. Rather than considering fault curves, we assume that every machine $u$ has a constant probability $p_u$ of failing. 
In this setting, there are $2^N$ possible combinations of machine failures (\textit{failure configurations}). 
Each configuration yields a set of possible system runs, which may differ based on the scheduling of messages. 
We deem a configuration safe if all of its system runs ensure agreement across non-failed nodes. We consider a configuration live if in all runs, all non-failed nodes eventually commit all operations.
By calculating how likely each failure configuration is, we can compute the overall probability that an algorithm guarantees safety and liveness in this specific deployment environment.

\subsection{Consensus Primer}\label{subsec:consensus-primer}

Most consensus protocols follow a similar structure. They proceed in a sequence of views, where each view is led by a distinct leader tasked with proposing client operations.
Within each view, committing an operation requires progressing through a series of steps, where one or more nodes broadcast a message and wait for a set of replies (a \textit{quorum}) before moving to the next step:
\par \textit{Step 1. Non-Equivocation (for BFT only).} This step ensures that, within a view, at most one operation will reach agreement per each slot. This phase is only necessary to defend against Byzantine leaders, as these can send conflicting proposals to nodes in the system. The non-equivocation phase terminates once participants obtain a non-equivocation quorum ($Q_{eq}$).
\par \textit{Step 2. Persistence.} This next step  ensures that any (possibly) committed operation is preserved across view and leader changes. Nodes commit operations once they obtain a persistence quorum ($Q_{per}$).
\par \textit{Step 3. View Change.}
When nodes detect that the consensus protocol is no longer making progress, a new leader is elected by receiving a view-change quorum ($Q_{vc}$). As spurious view changes can hamper liveness, most BFT consensus protocols  ensure that 
correct nodes join new views only after hearing about those views from sufficiently many other correct nodes (a view-change trigger quorum $Q_{vc\_t}$).

The size of each of these quorums depends on the invariants they wish to maintain and the failure models they assume. In BFT, non-equivocation quorums must intersect in at least one correct node to ensure that no two quorums can form for the same operation (a correct node will never vote for both). The view change quorum $Q_{vc}$ and persistence quorum $Q_{per}$ must also intersect in one correct node, thus ensuring that all committed operations will be included in the next view. To avoid spurious view changes, $Q_{vc\_t}$ must be guaranteed to include at least one correct node (correct nodes will not fabricate a view-change). 

The required invariants are simpler in the CFT setting: $Q_{per}$ need, for instance, only intersect $Q_{vc}$ in one correct node to ensure persistence across views. 
Violating any of these invariants will trigger a safety violation, while failing to form any of these quorums (for instance, because too many nodes have failed) will violate liveness.

\subsection{Analysis and Key Takeaways}
Specialising the aforementionned invariants for PBFT and Raft, two popular BFT and CFT protocols, yields the following two theorems, for a specific failure configuration. 

\begin{theorem}\label{thm:pbft}{\ }\\
PBFT is safe iff all these conditions hold:
\begin{enumerate}
 \item $|Byz| < 2|Q_{eq}| - N$ \hfill
 \item $|Byz| < |Q_{per}| + |Q_{vc}| - N$
\end{enumerate}
PBFT is live iff all these conditions hold:
\begin{enumerate}
  \item $|Byz| \le |Q_{vc\_t}| - |Q_{vc}|$
  \item $|Correct| \ge |Q_{eq}|,|Q_{per}|, |Q_{vc}|$ 
  \item $|Byz| < |Q_{vc\_t}|$
\end{enumerate}
\end{theorem}

Safety conditions (1,2) state that quorum sizes need to be large enough to ensure intersection in at least one correct node for a system of size N. Liveness instead requires that quorums be small enough that there will always be sufficiently many correct nodes to assemble said quorums. The $f$-threshold model simply counts the size of quorums to determine if they intersect, but a
more precise accounting is possible if we know the servers' fault probabilities.


\begin{table}
    \resizebox{\columnwidth}{!}{
    \begin{tabular}{|c|c|c|c|c|c|c|c|c|}
        \hline
        $N$ & $|Q_{eq}|$ & $|Q_{per}|$ & $|Q_{vc}|$ & $|Q_{vc\_t}|$ & Safe \% & Live \% & Safe and Live \% \\
        \hline
         4 & 3 & 3 & 3 & 2 & 99.94\% & 99.94\% & 99.94\% \\
         5 & 4 & 4 & 4 & 2 & 99.9990\% & 99.90\% & 99.90\% \\
         7 & 5 & 5 & 5 & 3 & 99.997\% & 99.997\% & 99.997\% \\
         8 & 6 & 6 & 6 & 3 & 99.99993\% & 99.995\% & 99.995\% \\
         \hline
    \end{tabular}
    }
    \caption{PBFT reliability, uniform $p_u=1\%$.}
    \label{tab:pbft}
\end{table}

\begin{theorem}\label{thm:raft}{\ }\\
Raft is safe iff all these conditions hold:
\begin{enumerate}
    \item $N < |Q_{per}| + |Q_{vc}|$ and \hfill
    \item $N < 2|Q_{vc}|$ \hfill
\end{enumerate}
    Raft is live iff:
\begin{enumerate}
    \item $|Correct| \ge |Q_{per}|, |Q_{vc}|$
\end{enumerate}
\end{theorem}
Safety conditions (1) and (2) state that quorums must be large enough for any two quorums to intersect in at least one node, ensuring (1) operations persist across views and (2) a unique leader is elected.

The probability of the algorithm being safe and/or live can then simply be calculated by summing the probability of each safe (respectively live) failure configuration. Exploring how fault probabilities impact safety/liveness guarantees across different network and quorum sizes yields several interesting observations (Table~\ref{tab:pbft} and ~\ref{tab:raft}).

\begin{table}
    \resizebox{\columnwidth}{!}{
    \begin{tabular}{|c|c|c|c|c|c|c|}
        \hline
        $N$ & $|Q_{per}|$ & $|Q_{vc}|$ & S\&L $p_u=1\%$ &  S\&L $p_u=2\%$ & S\&L $p_u=4\%$ & S\&L $p_u=8\%$ \\
        \hline
         3 & 2 & 2 & 99.97\% & 99.88\% & 99.53\% & 98.18\% \\
         5 & 3 & 3 & 99.9990\% & 99.992\% & 99.94\% & 99.55\% \\
         7 & 4 & 4 & 99.99997\% & 99.9995\% & 99.992\% & 99.88\% \\
         9 & 5 & 5 & 99.999998\% & 99.99996\% & 99.9988\% & 99.97\% \\
         \hline
    \end{tabular}
    }
\caption{Raft reliability for uniform node failure $p_u$.}
    \label{tab:raft}
\end{table}

\par 
\textbf{Consensus is probabilistic, like it or not.} $f$-threshold protocols assert that they are fully safe and live when $N=3$ and $f=1$. In reality, our analysis reveals that Raft with $N=3$ is only 3 nines safe and live  ($p_u=1\%$) (\autoref{tab:raft}).

\par \textbf{Linear size quorums can be overkill.}
Quorums in PBFT and Raft (\S\ref{subsec:consensus-primer}) grow linearly with the network size. For instance, an $Q_{vc\_t}$ quorum is at least $f+1$ in size ($N=100, |Q_{vc\_t}|=34$) to ensure that at least one correct node requests a view change~\cite{libra}. 
This is overkill: if $p_u=1\%$, there are already ten nines of probability that a random quorum of five nodes includes at least one correct node.

\par \textbf{Larger networks of less reliable nodes can help.}
 We find that a three-node Raft cluster ($p_u=1\%$) has equal safety/liveness probability as a nine node cluster with $p_u=8\%$. If reliability is proportional to pricing (\eg Spot instances), this could yield 3$\times$ lower cost. Hardware operators can thus use this analysis to pick the most sustainable, affordable, and/or performant hardware with no reliability trade-off. Operators could similary reuse older hardware to reduce carbon emissions 
 while meeting the target quality-of-service (QoS) for reliability~\cite{barroso2019datacenter,gupta2022act,wang2024designing,buffaloReducingCarbon,meza2015large}.
\noindent \textbf{Raft and PBFT underutilize reliable nodes.}
Raft and PBFT are oblivious to fault curves. Consider a seven node cluster with $p_u{=}8\%$ nodes running Raft. This cluster is $99.88\%$ safe (\autoref{tab:raft}). If we replace three nodes
with more reliable ones ($p_u{=}1\%$)---almost half of the nodes---safety improves only to
$99.98\%$ (not shown in table). As Raft 
does not know which nodes are more reliable, it may persist data only on the unreliable nodes.
If we required quorums to include at least one reliable node (by leveraging knowledge of fault curves), data durability would increase to $99.994\%$.

\noindent \textbf{There is a hidden exploitable trade-off between safety and liveness.}
The $f$-threshold model hides an inherent trade-off between safety and liveness in consensus protocols.
Exposing this trade-off can save resources.
Consider $f{=}1$ and two PBFT systems, one with $3f{+}1{=}4$ nodes and the other with $3f{+}2{=}5$ nodes.
In the $f$-threshold model, both systems tolerate $1$ fault, so these is no reason to deploy $5$ nodes.
However, in the probabilistic world, our analysis finds that using $5$ nodes improves PBFT safety by
42--60$\times$ with a small $1.67\times$ decrease in liveness compared to 4 nodes (\autoref{tab:pbft})---in fact, the 5-node
   system is more safe than a 7-node system, which is 40\% more expensive to
   deploy and operate.
The safety gain in the 5-node system over 4 nodes occurs because it has larger quorums.
Larger quorums improve safety (better probability of
  intersection) but degrade liveness (fewer failures can prevent progress)---in this case just a little.
The 7-node system, despite tolerating an additional failure ($f{=}2$), 
  increases the odds of faults due to its larger size, which partly offsets its gain in safety due to tolerating more faults.
The $f$-threshold model not only hides these insights, but also misleads.

\section{A Probabilistic Vision}
\label{sec:call}

The takeaways from our early analysis inspire several promising directions for designing probabilistic consensus; we first outline two challenges that must be addressed to realize usable probabilistic consensus: capturing an accurate fault model and reasoning about end-to-end guarantees.


\par \textbf{Accurate fault curves}. 
This research hinges on the ability to accurately express, in simple terms, potentially complex fault curves. Pessimistic characterizations will hurt performance while overly simplified or optimistic ones may cause the system to break when deployed.
Fault curves can be computed from telemetry, proactive monitoring for failures, studies modeling hardware faults, ... The storage community already relies on such data to model failure rates of disks~\cite{heart};
they rely on realistic estimates of failure probabilities and repair or disaster recovery probabilities, to design reliability mechanisms~\cite{patterson1988case}. 
These numbers are then used to derive metrics like MTBF or MTTDL, thus defining reliability as the expected time until "something bad happens". Consensus, in contrast, has always been designed to optimistically prevent "bad things" from ever happening. Future research should address this mismatch when formalizing fault curves.
Moreover, modeling correlated failures remains an open challenge; Markov models, for instance, which are typically used to compute MTBF and MTTDL are unable to capture dependent system transitions~\cite{tschaikowski2014tackling-markov-explosion,greenan2010mean}.

\par \textbf{End-to-end guarantees.} Applications care about end-to-end reliability guarantees, where consensus is a small part of the system. 
Traditional reliability guarantees~\cite{ford2010availability, cidon2013copysets}, expressed in terms of nines of \textit{availability} or \textit{durability}, do not align well with even the probabilistic type of safety and liveness offered by consensus. 
A consensus protocol that is $>0\%$ available will ensure the system remains live. A live consensus protocol, however, might not be able to meet the availability requirements if its recovery or reconfiguration is intolerably slow. Outside of availability, an unsafe system may commit different operations at different nodes yet remain durable if both forks are preserved.  

\noindent \textbf{Towards Probability-native consensus}. Once we have accurate fault curves, the next question becomes, how do we use them?  Our preliminary analysis suggests multiple steps.
First, we can incorporate fault curves and probabilistic safety/liveness into \textit{existing} consensus protocols. 
For instance, we can choose quorum sizes dynamically such that they overlap with high probability. Even this seemingly simple step is non-trivial as quorums are not formed independently, but instead must intersect~\cite{howard2016flexible}. This dependence makes calculating the probability of this intersection significantly more challenging as traditional tools like Chernoff bounds~\cite{malkhi1997probabilistic,p-pbft} no longer apply.

Second, probabilistic approaches can choose leaders among the most reliable nodes, avoiding more failure-prone nodes. This is similar in spirit to leader reputation schemes~\cite{cohen2021awareleaders, shoal} in the $f$-threshold model. Such a strategy can improve tail latency, reduce reconfiguration delays, and improve safety when nodes fail.
Probabilistic approaches can be further used to design new types of failure detectors~\cite{fail-detect1, fail-detect2}, which are more realistic and accurate.
Similarly, predictive models for node reliability enable preemptive reconfiguration, mitigating potential failures from jeopardizing safety or liveness. 

Third, in deployments where nodes' reliability exceeds application requirements, probabilistic protocols can sample committees, in particular, to select only the reliable nodes.

Finally, choosing to lean in fully into the probabilistic nature of consensus allows us to explore more radical design decisions. 
For instance, most consensus protocols have been designed around a few fundamental concepts such as majority-based quorum intersection. Probabilistic abstractions call for re-imagining consensus beyond quorums (\eg like in Ben-Or~\cite{ben-or} or Rabia~\cite{pan2021rabia}). The nature of quorum systems is, by definition, pessimistic: they guarantee that any two quorums will always intersect. In practice, however, sampling from much smaller subsets of nodes can guarantee intersection with high enough probability. Similarly, quorum systems that enforce durability are too conservative as they consider worst-case adversarial scenarios. In theory, they no longer guarantee safety if \textit{any} combination of $|Q_{per}|$ nodes fail. But, in reality, the probability that $|Q_{per}|$ failures leads to data loss is vanishingly unlikely.
For example, in a 100 node cluster where $|Q_{per}| = 10$ and $p_u=10\%$ there is a $50\%$ chance that $|Q_{per}|$ faults occur. However, for this situation to incur data loss, the $|Q_{per}|$ failures must perfectly overlap with the most recently formed persistence quorum which has a one in ten billion probability.

\section{Related Work}
\label{sec:related-work}


\textbf{Quorum Systems.} Prior works~\cite{byz-quorum-systems, quorum-systems-naor} introduce measures of load, capacity, and availability for quorums; however, they assume each node fails with equal probability.
Probabilistic quorums~\cite{malkhi1997probabilistic,abraham2003probabilistic,p-pbft} relax the traditional quorum intersection requirements with smaller, $O(\sqrt{N})$-sized quorums that overlap with high probability.



\noindent \textbf{Committee sampling.} King and Saia~\cite{king2009almost} propose a consensus mechanism that achieves $O(n^{1.5})$ communication complexity, by selecting subsets of the network contain a fraction of faulty servers representative of the entire cluster.
Algorand~\cite{algorand} leverages Verifiable Random Functions (VRFs) for efficient and secure random committee sampling.


\noindent \textbf{Refined failure and trust models.}
Upright~\cite{clement09upright} introduces separate thresholds for crash and Byzantine failures.
Stake-based consensus protocols~\cite{proof-of-stake, ganesh2019proof, gilad2017algorand} assume that servers have unique stake and more than $f$ stake will never fail simultaneously.
Stellar~\cite{mazieres2015stellar} generalizes this approach to enable collective agreement among servers with differing views on stake assignment.
While these more expressive failure models each address different shortcomings of the $f$-threshold model, they are unable to take full advantage of the rich knowledge of failures in practice to provide usable, end-to-end guarantees at low overheads.



\noindent \textbf{Analysis of $f$-threshold systems.} Zorfu~\cite{anderson2015recovery} uses Markov analysis to study mean time to $>f$ failures in $f$-threshold consensus systems. However, it does not extend this analysis to mean time to data loss (MTTDL) nor does it design its consensus algorithm based on failure rates.
~\cite{early-stopping, phase-king} analyze round complexity of $f$-threshold consensus protocols when the number of actual failures is less than $f$. ~\cite{bft2f, bft-forensics} analyze consistency and accountability when the number of actual failures exceed $f$ but are less than $2f$.

\section{Conclusion}
\label{sec:conc}

This paper argues that the $f$-threshold model is well-intended but ultimately unhelpful. Instead, it argues for explicitly capturing the probabilistic, evolving nature of hardware faults, much like the storage community already does. With this shift, we envision the emergence of new, more efficient consensus protocols that better align with how people think about reliability today.

\bibliographystyle{plain}
\bibliography{refs,references}

\end{document}